\documentclass{PoS}

\usepackage{graphicx}
\usepackage{amssymb}
\usepackage{graphicx}
\usepackage{dcolumn}
\usepackage{bm}
\usepackage{amsmath}

\newcommand{\Mpl}{M_\mathrm{Pl}}

\newcommand{\bmat}{\beta_\mathrm{m}}
\newcommand{\bgam}{\beta_\gamma}
\newcommand{\rhom}{\rho_\mathrm{m}}

\newcommand{\meff}{m_\mathrm{eff}}
\newcommand{\Pgc}{{\mathcal P_{\gamma \leftrightarrow \phi}}}

\newcommand{\edet}{\epsilon_\mathrm{det}} 

\newcommand{\fgam}{F_\gamma}
\newcommand{\faft}{F_\mathrm{aft}}

\newcommand{\Gdec}{\Gamma_{\mathrm{dec,}\gamma}}
\newcommand{\Gaft}{\Gamma_\mathrm{aft}}
\newcommand{\xiref}{\xi_\mathrm{ref}}
\newcommand{\xhat}{{\hat{x}}}

\newcommand{\khat}{{\hat{k}}}

\title{The CHASE laboratory search for chameleon dark energy}

\ShortTitle{The CHASE laboratory search for chameleon dark energy}

\author{\speaker{Jason H. STEFFEN}\\
Fermilab Center for Particle Astrophysics\\
        E-mail: \email{jsteffen@fnal.gov}}

\author{For the GammeV--CHASE collaboration\\        
        }

\abstract{A scalar field is a favorite candidate for the particle responsible for dark energy.  However, few theoretical means exist that can simultaneously explain the observed acceleration of the Universe and evade tests of gravity.  The chameleon mechanism, whereby the properties of a particle depend upon the local environment, is one possible avenue.  We present the results of the Chameleon Afterglow Search (CHASE) experiment, a laboratory probe for chameleon dark energy.  CHASE marks a significant improvement other searches for chameleons both in terms of its sensitivity to the photon/chameleon coupling as well as its sensitivity to the classes of chameleon dark energy models and standard power-law models.  Since chameleon dark energy is virtually indistinguishable from a cosmological constant, CHASE tests dark energy models in a manner not accessible to astronomical surveys.}

\FullConference{Identification of Dark Matter 2010-IDM2010\\
		July 26-30, 2010\\
		Montpellier France}

\begin{document}

\section{Introduction}

A variety of observational evidence indicates that the expansion of the universe is accelerating, for which a promising class of explanations is scalar field ``dark energy'' with negative pressure~\cite{Caldwell_Dave_Steinhardt_1998}.  Such a field is expected to couple to Standard Model particles with gravitational strength and would mediate a new ``fifth'' force, but such forces are excluded by experiments on a wide range of scales.  Three known ways to hide dark energy-mediated fifth forces include: weak or pseudoscalar couplings between dark energy and matter~\cite{Frieman_Hill_Stebbins_Waga_1995}; effectively weak couplings locally~\cite{Dvali_Gabadadze_Porrati_2000}; and an effectively large field mass locally, as in chameleon theories~\cite{Khoury_Weltman_2004a,Khoury_Weltman_2004b,Brax_etal_2004}.

Chameleons are scalar (or pseudoscalar) fields with a nonlinear potential and a coupling to the local energy density.  They evade fifth force constraints by increasing their effective mass in high-density environments, while remaining light in the intergalactic medium.  Gravity experiments in the lab~\cite{Adelberger_etal_2009} and in space~\cite{Khoury_Weltman_2004a,Khoury_Weltman_2004b} can exclude chameleons with gravitational strength matter couplings, but strongly coupled chameleons evade these constraints~\cite{Mota_Shaw_2006,Mota_Shaw_2007}.  Casimir force experiments rule out strongly coupled chameleons~\cite{Brax_etal_2007c}, but are ineffective for a large class of potentials commonly used to model dark energy.  Collider data exclude extremely strongly coupled chameleons~\cite{Brax_etal_2009}.

Photon-coupled chameleons may be detected through laser experiments~\cite{Chou_etal_2009} or excitations in radio frequency cavities~\cite{Rybka_etal_2010}.  In laser experiments, photons travelling through a vacuum chamber immersed in a magnetic field oscillate into chameleons.  They are then trapped through the chameleon mechanism by the dense walls and windows of the chamber since their higher effective mass within those materials creates an impenetrably large potential barrier~\cite{Chou_etal_2009,Ahlers_etal_2008,Gies_Mota_Shaw_2008}.  After a population of chameleons is produced, the laser is turned off and a photodetector exposed in order to observe the photon afterglow as trapped chameleons oscillate back to photons.  The original GammeV experiment included a search for this afterglow and set limits on photon/chameleon couplings below collider constraints for a limited set of dark energy models~\cite{Chou_etal_2009}.  The GammeV Chameleon Afterglow Search (CHASE) is a new experiment to search for photon coupled chameleons~\cite{Steffen_etal_2010}.  Its results bridge the gap between GammeV~\cite{Chou_etal_2009} and collider constraints, improves sensitivity to both matter and photon couplings to chameleons, and probes a broad class of chameleon models.

\section{Chameleon Models}

We consider actions of the form
\begin{equation}
S =\int d^4x \sqrt{-g}{\Big (} \frac{1}{2}\Mpl^2R
- \frac{1}{2}\partial_\mu\phi\partial^\mu\phi - V(\phi)
- \frac{1}{4}e^{\bgam \phi/\Mpl}F^{\mu\nu}F_{\mu\nu}
+ {\mathcal L}_\mathrm{m}(e^{2\bmat\phi/\Mpl}g_{\mu\nu},\psi^i_\mathrm{m}) {\Big )}
\label{e:action}
\end{equation}
where the reduced Planck mass $\Mpl = 2.43\times 10^{18}$~GeV, ${\mathcal L}_\mathrm{m}$ the Lagrangian for matter fields $\psi^i_\mathrm{m}$, and $\bgam$ and $\bmat$ are dimensionless chameleon couplings to photons and matter respectively (often expressed as $g_\gamma = \bgam / \Mpl$ and $g_\mathrm{m} = \bmat/\Mpl$).  We assume universal matter couplings.

The dynamics of this field are governed by an effective potential that depends on a potential $V(\phi)$, the background matter density $\rhom$, and the electromagnetic field Lagrangian density $\rho_\gamma = F^{\mu\nu}F_{\mu\nu}/4 = (B^2-E^2)/2$ (for pseudoscalars $\rho_\gamma = F^{\mu\nu}\tilde{F}_{\mu\nu}/4 = \vec{B}\cdot \vec{E}$):
\begin{equation}
V_{\rm eff}(\phi, \vec{x}) = V(\phi) + e^\frac{\bmat \phi}{\Mpl}\rhom(\vec{x}) +e^{\frac{\bgam \phi}{\Mpl} }\rho_\gamma(\vec{x}).
\end{equation}
A well-studied class of chameleon models has a potential of the form~\cite{Brax_etal_2004}
\begin{equation}
V(\phi)
=
M_\Lambda^4
e^{\kappa \left(\frac{\phi}{M_\Lambda}\right)^N}
\approx
M_\Lambda^4\left[  1 +  \kappa \left(\frac{\phi}{M_\Lambda}\right)^N \right].
\label{e:V_pwr}
\end{equation}
where $N$ is a real number and $M_\Lambda = \rho_\mathrm{de}^{1/4} \approx 2.4\times 10^{-3}$~eV is the mass scale of the dark energy density $\rho_\mathrm{de}$ and $\kappa$ is a dimensionless constant.  The constant term in this potential causes cosmic acceleration that is indistinguishable from a cosmological constant for cosmological surveys.  However, the local dynamics from the power-law term can be probed in the laboratory.

Following the derivations in \cite{Raffelt_Stodolsky_1988,Upadhye_Steffen_Weltman_2010} the conversion probability between photons and chameleons is
\begin{equation}
\Pgc
=
\left(\frac{2 \omega \bgam B}{\Mpl \meff^2}\right)^2
\sin^2\left(\frac{\meff^2 \ell}{4\omega}\right)
\khat \times (\xhat \times \khat).
\label{e:prob}
\end{equation}
Here, $\omega$ is the particle energy, $\meff = \sqrt{V_{\mathrm{eff,}\phi\phi}}$ 
is the effective chameleon mass in the environment, $\ell$ is the distance travelled through the magnetic field, and $\khat$ is the particle direction.

When a photon/chameleon wavefunction strikes an opaque surface of the vacuum chamber, there is a model-dependent phase shift $\xiref$ between the two components and a reduction in photon amplitude due to absorption.  On the other hand, a glass window performs a quantum measurement---chameleons reflect while photons are transmitted.  The velocities of trapped chameleons quickly become isotropic through surface imperfections.  The decay rate of a chameleon to a photon $\Gdec$, is found by averaging over initial directions and positions.  The observable afterglow rate per chameleon $\Gaft$ is found by averaging over those trajectories that allow a photon to reach the detector.  Once the geometry of an experiment is defined, these rates can be computed numerically~\cite{Upadhye_Steffen_Weltman_2010}.

A single parameter $\eta$ can be used to describe the chameleon effect.  If the chameleon mass in the chamber is dominated by the matter coupling, then $\meff \propto \rhom^\eta$ where $\eta = (N-2) / (2N-2)$~\cite{Upadhye_Steffen_Weltman_2010}.  The largest value of $\eta$ with integer $N$ is $\eta = 3/4$ for $N=-1$; $\phi^4$ theory ($N=4$), has $\eta = 1/3$.  We do not consider $0<N<2$ since their potentials are either unbounded from below or do not exhibit the chameleon effect.

\section{Apparatus}

\begin{figure}[ht]
\begin{center}
\includegraphics[width=0.8\textwidth]{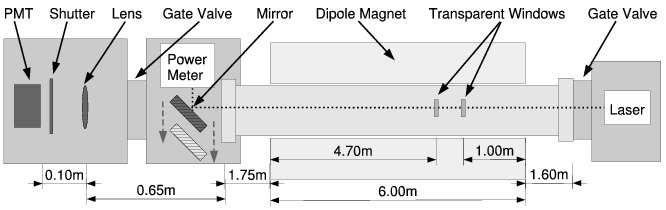}
\end{center}
\caption{
Schematic of the CHASE apparatus.\protect\label{schematic}
}
\end{figure}

The design of the CHASE apparatus is shown in Fig.~\ref{schematic}.  In addition to the windows at the ends of the vacuum chamber, we centered two glass windows in the cold bore of a Tevatron dipole magnet which divide the magnetic field into three partitions of lengths 1.0\,m, 0.3\,m, and 4.7\,m.  The shorter partition lengths provide sensitivity to larger-mass chameleons.

For a fixed magnetic field there are limits to the smallest and largest detectable $\bgam$---small $\bgam$ produce small afterglow signals while with large $\bgam$ the chameleon population will decay before the detector can be exposed.  We improve our sensitivity to large $\bgam$ by operating at a variety of lower magnetic fields, which lengthen the decay time of the chameleon population and provide overlapping regions of sensitivity.  A mechanical shutter (chopper) modulates any afterglow signal allowing a measurement of the PMT dark rate and improving sensitivity to low afterglow rates (small $\bgam$).

The CHASE vacuum system uses ion pumps and cryogenic pumping on the cold ($\sim 4$~K) bore of the magnet.  This design allows CHASE to probe $\eta$ as low as $0.1$ because of the large matter density ratio $\rhom(\mathrm{wall}) / \rhom(\mathrm{chamber}) \sim 10^{14}$ between the walls of the vacuum chamber and its interior and the large masses probed by the shortest partition of the magnetic field.

\section{Data}

We collected data in 14 configurations---seven magnetic field values (0.050, 0.090, 0.20, 0.45, 1.0, 2.2, and 5.0 Telsa) and both vertical and horizontal polarizations of the the laser.  We repeated measurements at 5.0T for a total of 16 science ``runs''.  A single science run consists of shining a frequency-doubled Nd:YAG laser through the cavity for ten minutes, then exposing the PMT to the apparatus for 14 minutes while cycling the shutter open and closed in $\sim 15$ second intervals each.  At the highest magnetic field the filling and observation stages are extended to 5 hours and 45 minutes respectively to improve our sensitivity to low $\bgam$.  The average input laser power is 3.5 Watts.  Before and after each science run, there is a 15 minute calibration run which we use to measure three quantities: the excess photon rate coming from the apparatus (due in part to discharge from the ion pumps), random fluctuations of that excess, and its run-to-run variation.  Our counting rate with the shutter-closed is $\sim 28$\,Hz, we see an excess rate of $1.15 \pm 0.08$\,Hz with the shutter open.  Our photon detection efficiency, including optical losses and the detection efficiencies of the PMT, is $\edet =  0.29$.

Following the filling stage, we observe a decaying rate of photons that we call ``orange glow''.  While its cause is unknown, its properties distinguish it from a scalar or pseudoscalar chameleon signal.  First, the orange glow is isolated to the red and orange parts, rather than the expected 532nm green.  Second, it is independent of both the strength of the magnetic field and the laser polarization.  Finally, the amplitude of the orange glow depends upon the temperature of the magnet bore.  Virtually disappearing at room temperature, it increases to several hundred Hz at temperatures near 4~K.  We eliminate the fast decaying components of the orange glow by ignoring the initial $\sim 120$ seconds of data from each science run.  This cut limits our sensitivity to large photon couplings.  However, data at lower magnetic fields compensates for this limitation.

\section{Analysis}

For the CHASE geometry, the rates $\Gdec$ and $\Gaft$ are computed in \cite{Upadhye_Steffen_Weltman_2010}.  We extend those calculations to greater $\meff$ using a Monte Carlo simulation.  We account for the absorption of photon amplitude and the differences in the induced phase shift $\xiref$ between the $s$ and $p$ polarizations at each reflection from the stainless steel surfaces.  Given these rates, the population $N_\phi$ of chameleons in the vacuum chamber is found by integrating
\begin{equation}
\frac{d N_\phi}{dt}
=
\fgam(t) \Pgc
-
N_\phi(t) \Gdec
\end{equation}
where $\fgam(t)$ is the rate that laser photons stream through the chamber.  Afterglow photons emerge and hit the PMT at a rate $\faft(t) = \edet N_\phi(t)\Gaft$.

For a given $\vec B$, each chameleon model---specified by its mass $\meff$ in the vacuum chamber, its photon coupling $\bgam$, and the chameleon potential $V(\phi)$---has a characteristic afterglow signal.  We compute this signal as a function of the magnetic field and the chameleon model parameters.  From the raw data binned to the $\sim 15$\,sec shutter cycle (e.g., Fig.~\ref{f:data}) we subtract the 1.15\,Hz excess rate from the ion pumps and the mean dark rate measured with the shutter-closed data for that run.  Statistical fluctuations including the observed excess are included in the uncertainty of each datum.  Data for all science runs with a given laser polarization and for a set of runs with $\vec{B} = 0$ are simultaneously analyzed using the Profile Likelihood method~\cite{rolk2005}.  As nuisance parameters we include a common, exponentially decaying signal, which eliminates the long-decay tail of the orange glow, and each run is allowed an independent constant offset constrained by the possible 0.40\,Hz run-to-run variations in the ion pump glow.  We compare the $\chi^2$ for the chameleon model with that for the model where no chameleon is present.  Any chameleon model whose $\chi^2$ is greater by $6.0$ is excluded to $95\%$ confidence.


\section{Results}

\begin{figure}[tb]
\begin{center}
\includegraphics[width=2.7in]{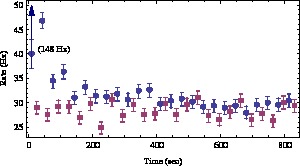} \quad
\includegraphics[width=0.48\textwidth]{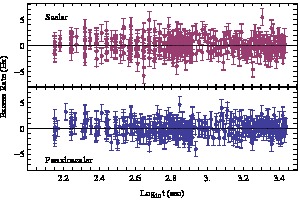}
\caption{Left: Example of CHASE raw data given as the observed PMT rate vs. time.  Blue circles and red squares are data with the shutter open and closed respectively.  The left most datum would appear off the scale at 148Hz.  Right: Residuals from the null model for all CHASE science data.  The lower panel is for pseudoscalar and the upper for scalar chameleons.  Data for all magnetic fields are overlaid.\label{f:data}}
\end{center}
\end{figure}

\begin{figure}[tb]
\begin{center}
\includegraphics[angle=270,width=2.8in]{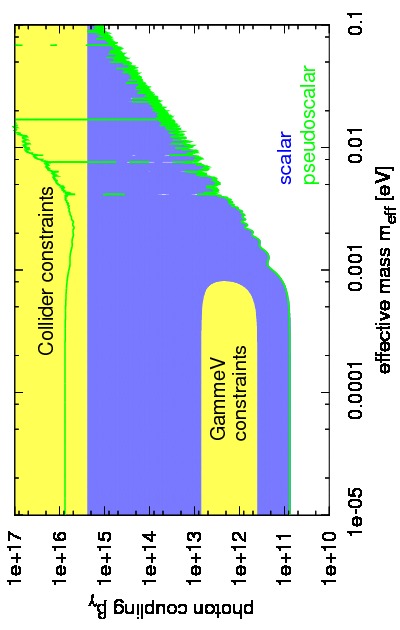}
\includegraphics[angle=270,width=2.8in]{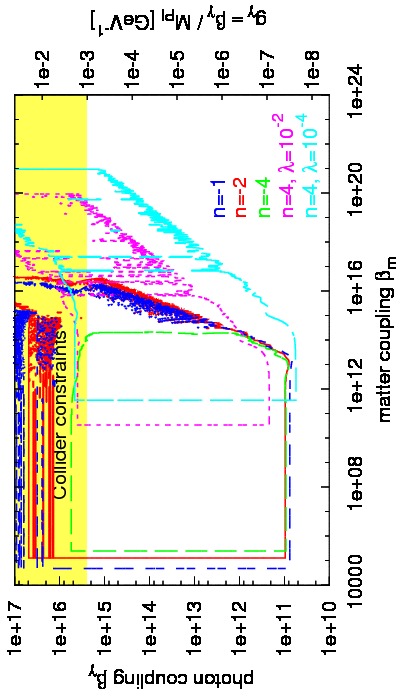}
\includegraphics[width=0.75\textwidth]{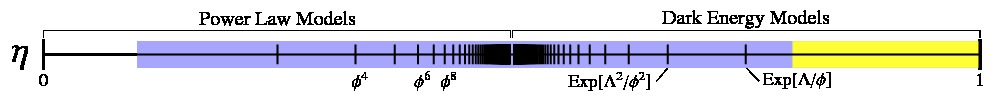}
\caption{Left: Scalar (solid) and pseudoscalar (outline) constraints, at $95\%$ confidence, in the ($\meff$, $\bgam$) plane for $\xiref = 0$.  Right: $95\%$ confidence-level constraints on chameleons with power law potentials (2.3).  For potentials whith $N<0$ we set $\kappa = 1$; for $\phi^4$ theory ($N=4$), we use the standard $\kappa = \lambda/4!$.  Bottom: Chameleon models probed by CHASE as parameterized by $\eta$.  GammeV sensitivity is yellow while CHASE sensitivity is blue.\label{f:constraints_model-indep}}
\end{center}
\end{figure}

Analysis of our data shows no evidence for a photon-coupled chameleon.  Figure \ref{f:data} shows all of the residuals in the science data for the no chameleon model.  The mean and RMS of these residuals are $0.05$ and $1.35$\,Hz for pseudoscalar couplings ($\chi^2 = 421$ with 471 degrees of freedom (DOF)) and $0.06$ and $1.62$\,Hz for scalar couplings ($\chi^2 = 502$ with 472 DOF).  Fig.~\ref{f:constraints_model-indep} shows parameters excluded to $95\%$ confidence for scalars and pseudoscalars assuming $\meff$ dependence on $B$ to be negligible and $\xiref = 0$.  These constraints reach four significant milestones.  First, they bridge the nearly three order of magnitude gap between bounds on $\bgam$ from GammeV and from colliders~\cite{Brax_etal_2009}.  Second, they exclude a range of $\bgam$ spanning four orders of magnitude at masses around the dark energy scale ($2.4 \times 10^{-3}$\,eV).  Third, they rule out photon couplings roughly an order of magnitude below previous limits in this mass range where $\bgam < 7.1\times 10^{10}$ for scalar and $\bgam < 7.6\times 10^{10}$ for pseudoscalar chameleons.  Finally, they are sensitive to chameleon dark energy models and chameleon power-law models where $\eta > 0.1$, including $V \propto \phi^4$.

Figure~\ref{f:constraints_model-indep} shows CHASE constraints (at 95\%) for select potentials given by Eq.~(\ref{e:V_pwr}).  These limits truncate at low $\bmat$ by the requirement that chameleons reflect from the chamber walls, at high $\bmat$ by destructive interference at large $\meff$ (see Fig.~\ref{f:constraints_model-indep}), and at low $\bgam$ by undetectably small signals.  Not surprisingly, theories with the largest $\eta$ are excluded over the greatest range of $\bmat$.  These constraints complement those from torsion pendula, which probe $\bmat \sim 1$, and are consistent with constraints from Casimir force measurements for $N=4$~\cite{Brax_etal_2007c}.  CHASE data exclude chameleons spanning five orders of magnitude in photon coupling and over 12 orders of magnitude in matter coupling for individual models.  They probe a wide range of chameleon models, and give significantly improved constraints for cosmologially interesting chameleon dark energy models.

\paragraph{Acknowledgements:} We thank the staff of the Fermilab Technical Division Test and Instrumentation Department, the Fermilab Particle Physics Division mechanical design and electrical engineering groups, and the vacuum experts of the Fermilab Accelerator Division.  This work is supported by the U.S. Department of Energy under contract No. DE-AC02-07CH11359 as well as by the Kavli Institute for Cosmological Physics under NSF contract PHY-0114422.

\end{document}